\documentclass[journal,twoside]{IEEEtran}
\usepackage{graphicx,cite,amssymb,amsmath,psfrag,subfigure}
\usepackage{graphicx}
\usepackage[displaymath,mathlines]{lineno}
\usepackage{cite}
\usepackage{subfigure}
\usepackage{mathrsfs}
\usepackage{color}
\usepackage{tabulary}
\usepackage{multirow}

\newtheorem{proposition}{Proposition}

\DeclareMathOperator{\E}{\mathbb{E}}

\newcommand{\CG}[2]{\mathcal{CN}\left({#1},{#2}\right)}

\newcommand{\B}[1]{{\mathbf{#1}}}

\newcommand{\Pu}{p_{\mathrm{u}}}
\newcommand{\Pp}{p_{\mathrm{p}}}

\begin{document}

\title{
     \huge{Massive MIMO with Optimal Power and Training Duration Allocation}}
\author{
        Hien Quoc Ngo, Michail~Matthaiou,  and
        Erik G. Larsson
\thanks{
        H.~Q.\ Ngo and E.~G.\ Larsson are with the Department of Electrical
    Engineering (ISY), Link\"{o}ping University, 581 83 Link\"{o}ping,
    Sweden
        (email: nqhien@isy.liu.se; egl@isy.liu.se).
}
\thanks{
M. Matthaiou is with the School of Electronics, Electrical
Engineering and Computer Science, Queen's University Belfast,
Belfast, BT3 9DT, U.K., and with the Department of Signals and
Systems, Chalmers University of Technology, 412 96 Gothenburg,
Sweden (email: m.matthaiou@qub.ac.uk). }
 }

\markboth{}
        {Massive MIMO with Optimal Power and Training Duration Allocation}

\maketitle

%\renewcommand{\baselinestretch}{1.5} \normalsize

%%%%%%%%%%%%%%%%%%%%%%%%%%%%%%%%%%%%%%%%%%%%%%%%%%%%%%%%%%%%%%%%%%%%%

\begin{abstract}
We consider the uplink of massive multicell multiple-input
multiple-output systems, where the base stations (BSs),  equipped
with massive arrays, serve simultaneously several terminals in the
same frequency band. We assume that the BS estimates the channel
from uplink training, and then uses the maximum ratio combining
technique to detect the signals transmitted from all terminals in
its own cell. We propose an optimal resource allocation scheme
which jointly selects the training duration, training signal
power, and data signal power in order to maximize the sum spectral
efficiency, for a given total energy budget spent in a coherence
interval. Numerical results verify the benefits of the optimal
resource allocation scheme. Furthermore, we show that more
training signal power should be used at low signal-to-noise ratio
(SNRs), and vice versa at high SNRs. Interestingly, for the entire
SNR regime, the optimal training duration is equal to the number
of terminals. \vspace{-0.3cm}

\end{abstract}

%\begin{keywords}
%Maximum ratio combining,  Massive MIMO, very large MIMO systems.
%\end{keywords}

%\IEEEpeerreviewmaketitle

\section{Introduction}
Massive multiple-input multiple-output (MIMO) has attracted a lot
of research interest recently
\cite{Mar:10:WCOM,LTEM:14:CM,YGFL:12:JSAC,TH:13:JSCN}. Typically,
the uplink transmission in massive MIMO systems consists of two
phases: uplink training (to estimate the channels) and uplink
payload data transmission. In previous works on massive MIMO, the transmit power of each symbol
is assumed to be the same during the training and data
transmission phases \cite{Mar:10:WCOM,NLM:13:TCOM}. However, this
equal power allocation policy causes a ``squaring effect'' in the
low power regime \cite{HH:03:IT}. The squaring effect comes from
the fact that when the transmit power is reduced, both the data
signal and the pilot signal suffer from a power reduction. As a
result, in the low power regime, the capacity scales as $\Pu^2$,
where $\Pu$ is the transmit power.

In this paper, we consider the uplink of  massive multicell MIMO
with maximum ratio combining (MRC) receivers at the base station
(BS). We consider MRC receivers since they are simple and perform
rather well in massive MIMO, particularly when the inherent effect
of channel estimation on intercell interference is taken into
account \cite{NLM:13:TCOM}.
Contrary to most prior works, we assume that the average transmit
powers of pilot symbol and  data
symbol are different. We investigate a resource allocation problem
which finds the transmit pilot power, transmit data power,
as well as, the training duration that maximize the sum spectral
efficiency for a given
 total energy budget spent in a coherence interval.  Our numerical results show appreciable
benefits of the proposed optimal resource allocation. At low
signal-to-noise ratios (SNRs), more power is needed for training to
reduce the squaring effect, while at high SNRs, more power is
allocated to data transmission.

Regarding related works,
\cite{HH:03:IT, RHS:07:JSASP,MUS:07:JSAC} elaborated on
a similar issue. In \cite{HH:03:IT, RHS:07:JSASP}, the authors
considered point-to-point MIMO systems, and in \cite{MUS:07:JSAC},
the authors considered single-input multiple-output multiple
access channels with scheduling. Most importantly, the performance
metric used in \cite{HH:03:IT, RHS:07:JSASP,MUS:07:JSAC} was the
mutual information without any specific signal processing. In this
work, however, we consider massive
multicell multiuser MIMO systems with MRC receivers and
demonstrate the strong potential of these configurations.

\section{Massive Multicell MIMO System Model} \label{sec: system}

We consider the uplink multicell MIMO system described in
\cite{NLM:13:TCOM}. The system has $L$ cells. Each cell includes
one $N$-antenna BS, and $K$ single-antenna
terminals, where $N\gg K$. All $L$ cells share the same frequency
band. The transmission comprises two phases: uplink training and
data transmission.\vspace{-0.3cm}

\subsection{Uplink Training}
In the uplink training phase, the BS estimates the channel from
received pilot signals transmitted from all terminals. In each
cell, $K$ terminals are assigned $K$ orthogonal pilot sequences of
length $\tau$ symbols ($K\leq \tau\leq T$), where $T$ is the
length of the coherence interval. Since the coherence interval is
limited, we assume that the same orthogonal pilot sequences are
reused in all $L$ cells. This causes the so-called \emph{pilot
contamination} \cite{Mar:10:WCOM}. Note that interference from data symbols is as bad
as interference from pilots \cite{NLM:13:TCOM}.

We denote by $\B{G}_{\ell i} \in \mathbb{C}^{N\times K}$ the channel
matrix between the BS in the $\ell$th cell and  the $K$ terminals in the
$i$th cell. The $(m,k)$th element of $\B{G}_{\ell i}$ is modeled as
\begin{align}\label{eq CM 1}
    g_{\ell imk}
    =
        h_{\ell imk} \sqrt{\beta_{\ell ik}},~~~ m=1, 2, ..., N,
\end{align}
where $h_{\ell imk} \sim \CG{0}{1}$ represents the small-scale fading
coefficient from the $m$th antenna of the $\ell$th BS to the $k$th
terminal in the $i$th cell, and $\sqrt{\beta_{\ell ik}}$ is a constant
that represents large-scale fading (pathloss and shadow fading).

At the $\ell$th BS, the minimum mean-square error channel estimate
for the $k$th column of the channel matrix $\B{G}_{\ell \ell}$ is
\cite{NLM:13:TCOM}
\begin{align} \label{eq CE MMSE 1}
    \hat{\B{g}}_{\ell \ell k}
    &\!=\!
        {\beta_{\ell \ell k}}\!\!
        \left(
        \sum_{j = 1}^{L}
            \beta_{\ell jk}
         +
        \frac{1}{\tau p_{\mathrm{p}}}
         \!\right)^{-1}\!\!
        \left(
        \sum_{j =1}^{L}
            \B{g}_{\ell jk}
         +
        \frac{{\B{w}}_{\ell k}}{\sqrt{\tau  p_{\mathrm{p}}}}
         \!\right),
\end{align}
where  $\Pp$ is the transmit power of each pilot symbol, and
$\B{w}_{\ell k} \sim \CG{\B{0}}{\B{I}_N}$ represents  additive noise.

\subsection{Data Transmission}
In this phase, all $K$ terminals send their data to the BS. Let $\sqrt{p_{\mathrm{u}}} \B{x}_i \in
\mathbb{C}^{K\times 1}$ be a vector of symbols transmitted from
the $K$ terminals in the $i$th cell, where $\E\left\{\B{x}_i
\B{x}_i^H \right\} = \B{I}_K$, $\E\{\cdot\}$ denotes expectation,  and $\Pu$ be the average
transmitted power of each terminal. The $N \times 1$ received
vector at the $\ell$th BS is given by
\begin{align} \label{eq MU-MIMO 1}
    \B{y}_\ell
    =
        \sqrt{p_{\mathrm{u}}}
        \sum_{i=1}^{L}
            \B{G}_{\ell i}
            \B{x}_i
        +
        \B{n}_\ell,
\end{align}
where $\B{n}_\ell \in \mathbb{C}^{N\times 1}$ is the  AWGN vector,
distributed as $\B{n}_\ell \sim \CG{\B{0}}{\B{I}_N}$. Then, BS $\ell$
uses MRC together with the channel estimate to detect the signals
transmitted from the $K$ terminals in its own cell. More
precisely, to detect the signal transmitted from the $k$th
terminal, $x_{\ell k}$, the received vector $\B{y}_\ell$ is
pre-multiplied with $\hat{\B{g}}_{\ell \ell k}^H$ to obtain:
\begin{align} \label{eq MU-MIMO MRC1}
    r_k
    \triangleq \hat{\B{g}}_{\ell \ell k}^H \B{y}_\ell
    &=
       {\sqrt{p_{\mathrm{u}}}
            \hat{\B{g}}_{\ell \ell k}^H
            \B{g}_{llk}
            x_{\ell k}}
       +
       {\sqrt{p_{\mathrm{u}}}\sum_{j\neq k}^K \hat{\B{g}}_{\ell \ell k}^H
            \B{g}_{\ell \ell j}
            x_{\ell j}}
            \nonumber
            \\
        &+
        {\sqrt{p_{\mathrm{u}}}
        \sum_{i\neq \ell}^{L}
            \hat{\B{g}}_{\ell \ell k}^H
            \B{G}_{\ell i}
            \B{x}_i}
        +
        {\hat{\B{g}}_{\ell \ell k}^H\B{n}_\ell},
\end{align}
and then $x_{\ell k}$ can be extracted directly from $r_k$.

\subsection{Sum Spectral Efficiency}\label{Sec Spect}
In our analysis, the performance metric is the sum spectral
efficiency (in bits/s/Hz). From \eqref{eq MU-MIMO MRC1}, and
following a similar methodology as in \cite{NLM:13:TCOM}, we
obtain an achievable ergodic rate of the transmission from the
$k$th terminal in the $\ell$th cell to its BS as:\footnote{The
achievable ergodic rate for the case of $\beta_{\ell \ell k}=1$ and
$\beta_{\ell ik}=\beta$ ($i\neq \ell$), for all $k$, was derived in
\cite{NLM:13:TCOM}, see Eq.~(73).}
\begin{align} \label{eq: SE MRC 1}
R_{\ell k}
    =
        \log_2
        \left(
            1
            +
            \frac{
                a_k \tau \Pp\Pu
                }{
                b_k \tau \Pp\Pu + c_k \Pu + d_k \tau \Pp +1
                }
        \right),
\end{align}
where $a_k
    \triangleq
    \beta_{\ell \ell k}^2 \left(N-1\right)$,
\begin{align*}
b_k
    &\triangleq
    \left(N-1\right)
    \sum_{i\neq \ell}^{L} \beta_{\ell ik}^2
    -
    \sum_{i=1}^{L} \beta_{\ell ik}^2
    +
        \left(\sum_{i=1}^{L} \sum_{j=1}^K \beta_{\ell ij}\right)
        \sum_{i=1}^{L} \beta_{\ell ik},\\
c_k
    &\triangleq
    \sum_{i=1}^{L} \sum_{j=1}^K \beta_{\ell ij}, ~ \text{and} ~
d_k
    \triangleq
    \sum_{i=1}^{L} \beta_{\ell ik}.
\end{align*}
The sum spectral efficiency is defined as
\begin{align} \label{eq: low SE 1}
    {\mathcal{S}}
    &\triangleq
        \left(1-\frac{\tau}{T} \right)
        \sum_{k=1}^K
        R_{\ell k}.
\end{align}

For $\Pu\ll 1$, and for $\Pp$ fixed regardless of $\Pu$, we have
\begin{align} \label{eq: low SE 1aa}
    {\mathcal{S}}
    =
    \log_2 e\left(1-\frac{\tau}{T} \right) \sum_{k=1}^K \frac{a_k \tau \Pp
                }{
                d_k \tau \Pp +1
                }\Pu + \mathcal{O}\left(\Pu^2\right),
\end{align}
while for $\Pp=\Pu$ (the choice considered in \cite{NLM:13:TCOM}
and other literature we are aware of), we have
\begin{align} \label{eq: low SE 1aaa}
    {\mathcal{S}}
    =
    \log_2 e\left(1-\frac{\tau}{T} \right) \sum_{k=1}^K a_k \tau \Pu^2 +
    \mathcal{O}\left(\Pu^3\right).
\end{align}

Interestingly, at low $\Pu$, the sum spectral efficiency scales
linearly with $N$ [since $a_k =
\beta_{\ell \ell k}^2(N-1)$], even though the number of unknown channel
parameters increases. We can see that for the case of $\Pp$ being
fixed regardless of $\Pu$, at $\Pu\ll 1$, the sum spectral
efficiency scales as $\Pu$. However, for the case of $\Pp = \Pu$,
at $\Pu \ll 1$, the sum spectral efficiency scales as $\Pu^2$. The
reason is that when $\Pu$ decreases and, hence, $\Pp$ decreases,
the quality of the channel estimate deteriorates, which leads to a
``squaring effect'' on the sum spectral efficiency
\cite{HH:03:IT}.

Consider now the bit energy of a system defined as the transmit
power expended  divided by the sum spectral efficiency:
\begin{align} \label{eq: EE 1}
    \eta
    &\triangleq
        \frac{\frac{\tau}{T}\Pp + \left(1-\frac{\tau}{T}\right)\Pu}{ {\mathcal{S}} }.
\end{align}
If $\Pp = \Pu$ as in previous
works, we have $\eta=\frac{\Pu}{ {\mathcal{S}} }$. Then, from
\eqref{eq: low SE 1aaa}, when the transmit power is reduced below
a certain threshold, the bit energy increases even when we reduce
the power (and, hence, reduce the spectral efficiency). As a
result, the minimum bit energy is achieved at a non-zero sum
spectral efficiency. Evidently, it is inefficient to operate below
this sum spectral efficiency. However, we can operate in this
regime if we use a large enough transmit power for uplink pilots,
and reduce the transmit power of data. This observation is clearly
outlined in the next section.

\section{Optimal Resource Allocation} \label{Sec:PA}
Using different powers for the
uplink training and data transmission phases improves the system
performance, especially in the wideband regime, where the spectral
efficiency is conventionally parameterized as an affine function
of the energy per bit \cite{Verdu:02:IT}. Motivated by this
observation, we consider a fundamental resource allocation
problem, which adjusts the data power, pilot power, and duration
of pilot sequences, to maximize the sum spectral efficiency given
in \eqref{eq: low SE 1}. Note
that, this resource allocation can be implemented at the BS.

Let $P$ be the  total transmit energy constraint for each terminal
in a coherence interval. Then, we have
\begin{align} \label{eq: PA 1}
    \tau \Pp + (T-\tau)\Pu
    \leq
    P.
\end{align}
When $\tau \Pp$ decreases, we can see from \eqref{eq CE MMSE 1}
that the effect of noise on the channel estimate escalates, and
hence the channel estimate degrades. However, under the total
energy constraint \eqref{eq: PA 1}, $(T-\tau)\Pu$ will increase,
and hence the system performance may improve. Conversely, we could
increase the accuracy of the channel estimate by using more power
for training. At the same time, we have to reduce the transmit
power for the data transmission phase to satisfy  \eqref{eq: PA
1}. Thus, there are optimal values of $\tau$, $\Pp$, and $\Pu$ which
maximize the sum spectral
efficiency for given  $P$ and $T$.

Once the total transmit energy per coherence interval and the
number of terminals are set, one can adjust the duration of pilot
sequences and the transmitted powers of pilots and data to
maximize the sum spectral efficiency. More precisely,
\begin{align} \label{eq P3 1}
    \mathcal{P}_1 : \left\{%
\begin{array}{l}
  \mathop {\max}\limits_{\Pu, \Pp, \tau} ~ \mathcal{S}\\
  \hspace{0.4cm} \text{s.t.} ~ ~ \tau \Pp \!+\! (T\!-\!\tau)\Pu \!=\! P\\
  \hspace{1.1cm} \Pp \geq 0, \Pu \geq 0 \\
  \hspace{1.1cm}  K \leq \tau \leq T, \left(\tau \in \mathbb{N} \right)\\
\end{array}%
\right.
\end{align}
where the inequality of the total energy constraint in \eqref{eq:
PA 1} becomes the equality in \eqref{eq P3 1}, due to the fact
that for a given $\tau$ and $\Pp$, $\mathcal{S}$ is an increasing
function of $\Pu$, and for a given $\tau$ and $\Pu$, $\mathcal{S}$
is an increasing function of $\Pp$. Hence, $\mathcal{S}$ is
maximized when $\tau \Pp + (T-\tau)\Pu = P$.

\begin{figure}[t]
    \centerline{\includegraphics[width=0.45\textwidth]{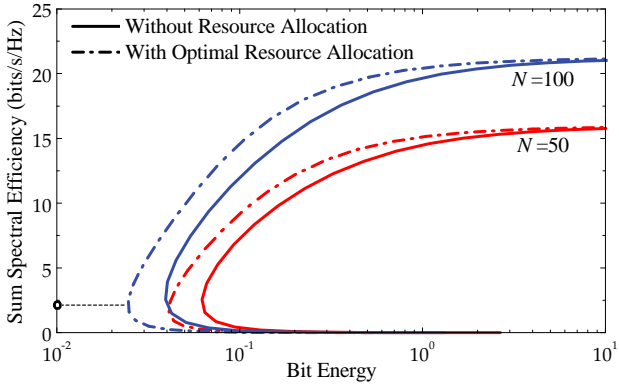}}
    \caption{Bit energy versus sum spectral efficiency with and without resource allocation.}
    \label{fig:4}
\end{figure}

\begin{figure}[t]
    \centerline{\includegraphics[width=0.45\textwidth]{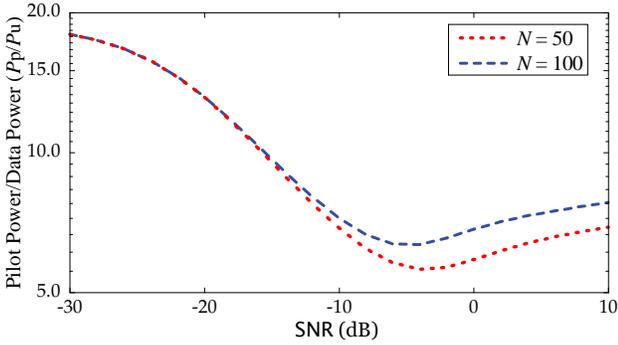}}
    \caption{Ratio of the transmit pilot power to the transmit data power.}
    \label{fig:5}
\end{figure}

\begin{proposition}\label{lemma1}
The optimal pilot duration, $\tau$, of $\mathcal{P}_1$ is equal to
the number of terminals $K$.
\begin{proof}
Let $\left(\tau^\ast, \Pp^\ast, \Pu^\ast\right)$ be a solution of
$\mathcal{P}_1$. Assume that $\tau^\ast>K$. Next we choose
$\bar{\tau}=K$, $\bar{p}_{\mathrm{p}}=\tau^\ast \Pp^\ast/K$, and
$\bar{p}_{\mathrm{u}}=\frac{P-\tau^\ast\Pp^\ast}{T-K}$. Clearly,
this choice of system parameters $\left(\bar{\tau},
\bar{p}_{\mathrm{p}}, \bar{p}_{\mathrm{u}}\right)$ satisfies the
constraints in \eqref{eq P3 1}. From \eqref{eq: low SE 1} and
using the fact that $\bar{\tau}\bar{p}_{\mathrm{p}}=\tau^\ast
\Pp^\ast$, we have $\mathcal{S}\left(\bar{\tau},
\bar{p}_{\mathrm{p}}, \bar{p}_{\mathrm{u}}\right)>
\mathcal{S}\left(\tau^\ast, \Pp^\ast, \Pu^\ast\right)$ which
contradicts the assumption. Therefore, $\tau^\ast=K$.
\end{proof}
\end{proposition}

From Proposition~\ref{lemma1}, $\mathcal{P}_1$ is equivalent to
the following optimization problem:
\begin{align}\label{Prob2}
    \mathcal{P}_2 : \left\{%
\begin{array}{l}
  \mathop {\max}\limits_{\Pu} ~\mathcal{S}|_{\Pp = P/K - (T/K -1)\Pu} \\
  \hspace{0.3cm} \text{s.t.} ~ ~ 0 \leq \Pu \leq  \frac{P}{T-K}.\\
\end{array}%
\right.
\end{align}

We can efficiently solve $\mathcal{P}_2$ based on the following
property:

\begin{proposition}\label{lemma 1}
The program $\mathcal{P}_2$ is concave.
\begin{proof}
See Appendix~\ref{proof Lemm1}.
\end{proof}
\end{proposition}

To solve the optimization problem $\mathcal{P}_2$, we can use any
nonlinear or convex optimization method to get the globally
optimal result. Here, we use the FMINCON function in MATLAB's
optimization toolbox.

\begin{figure}[t]
    \centerline{\includegraphics[width=0.45\textwidth]{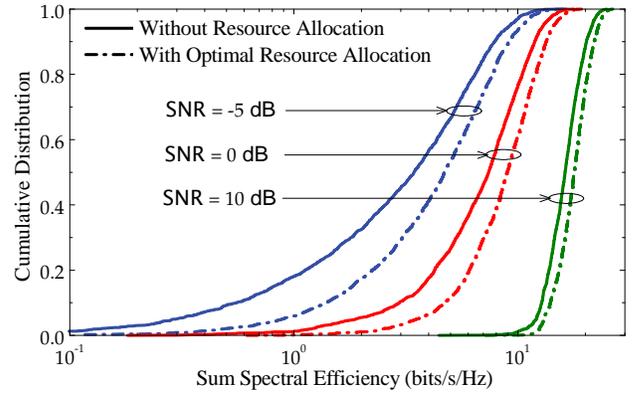}}
    \caption{Sum spectral efficiency with and without resource allocation ($N=100$).}
    \label{fig:6}
\end{figure}

\vspace{-0.5cm}
\section{Numerical Results} \label{sec
NR}

We consider a cellular network with $L=7$  hexagonal cells which
have a radius of $r_c=1000$m. Each cell serves $10$ terminals
($K=10$). We choose $T=200$, corresponding to a coherence
bandwidth of $200$~KHz and a coherence time of $1$~ms. We consider
the performance in the cell in the center of the network. We assume
that terminals are located uniformly and randomly in each cell and
no terminal is closer to the BS than $r_h=200$m. Large-scale
fading is modeled as $\beta_{\ell ik}=z_{\ell ik}/(r_{\ell ik}/r_h)^\nu$,
where $z_{\ell ik}$ is a log-normal random variable, $r_{\ell ik}$ denotes
the distance between the $k$th terminal in the $i$th cell and the
$\ell$th BS, and $\nu$ is the path loss exponent. We set the standard
deviation of $z_{\ell ik}$ to $8$dB, and $\nu=3.8$.

Firstly, we will examine the sum spectral efficiency versus the
bit energy obtained from one snapshot generated by the above
large-scale fading model. The bit energy is defined in \eqref{eq:
EE 1}. From \eqref{eq: EE 1} and \eqref{eq P3 1}, we can see that
the solution of $\mathcal{P}_1$ also leads to the minimum value of
the bit energy. Figure~\ref{fig:4} presents the sum spectral
efficiency versus the bit energy with optimal resource allocation.
As discussed in Section~\ref{Sec Spect}, the minimum bit energy is
achieved at a non-zero spectral efficiency. For example, with
optimal resource allocation, at $N=100$, the minimum bit energy is
achieved at a sum spectral efficiency of $2$ bits/s/Hz which is
marked by a circle in the figure. Below this value, the bit energy
increases as the sum spectral efficiency decreases. For a given
energy per bit, there are two operating points. Operating below
the sum spectral efficiency, at which the minimum energy per bit
is obtained, should be avoided.

On a different note, we can see that with optimal resource
allocation, the system performance improves significantly. For
example, to achieve the same sum spectral efficiency of $10$
bits/s/Hz, optimal resource allocation can improve the energy
efficiencies by factors of $1.45$ and $1.5$ compared to the case
of no resource allocation with $N=50$ and $N=100$, respectively.
This dramatic increase underscores the importance of resource
allocation in massive MIMO.
However, at high bit energy, the
squaring effect for the case of no resource allocation disappears
and, hence, the advantages of resource allocation diminish.
Furthermore, for the same  sum spectral efficiency,
$\mathcal{S}=10$ bits/s/Hz, and with resource allocation, by
doubling the number of BS antennas from $50$ to $100$, we can
improve the energy efficiency by a factor of $2.2$.

The corresponding ratio of the optimal pilot power to the optimal
transmitted data power for $N=50$ and $N=100$ is shown in
Fig.~\ref{fig:5}. Here, we define $\mathsf{SNR} \triangleq P/T$.
Since $P$ is the total transmit energy spent in a coherence
interval $T$ and the noise variance is $1$, $\mathsf{SNR}$ has the
interpretation of average transmit SNR and is therefore
dimensionless. We can see that at low $\mathsf{SNR}$ (or low
spectral efficiency), we spend more power during the training
phase, and vice versa at high $\mathsf{SNR}$. At low
$\mathsf{SNR}$, $\Pp/\Pu \approx 18$ which leads to $\tau \Pp /
{(T-\tau) \Pu} \approx 1$. This means that half of the total
energy is used for uplink training and the other half is used for
data transmission. Note that the
power allocation problem in the low SNR regime is useful since the
achievable rate (obtained under the assumption that the estimation
error is additive Gaussian noise) is very tight, due to the use of
Jensen's bound in \cite{NLM:13:TCOM}.
 Furthermore, in
general, the ratio of the optimal pilot power to the optimal
 data power does not always monotonically decrease with
increasing $\mathsf{SNR}$. We can see from the figure that, when
$\mathsf{SNR}$ is around $-5$dB, $\Pp/\Pu$ increases
when $\mathsf{SNR}$ increases.

We now consider the cumulative
distribution of the sum spectral efficiency obtained from $2000$
snapshots of large-scale fading (c.f. Fig.~\ref{fig:6}). As
expected, our resource allocation improves the system performance
substantially, especially at low SNR. More importantly, with
resource allocation, the sum spectral efficiencies are more
concentrated around their means compared to the case of no
resource allocation. For example, at $\mathsf{SNR}=0$dB,  resource
allocation increases the $0.95$-likely sum spectral efficiency by
a factor of $2$ compared to the case of no resource allocation.

\section{Conclusion} \label{Sec:Conclusion}

Conventionally, in massive MIMO, the transmit powers of the pilot
signal and data payload signal are assumed to be equal. In this
paper, we have posed and answered a basic question about the
operation of massive MIMO: How much would the performance improve
if the relative energy of the pilot waveform, compared to that of
the payload waveform, were chosen optimally? The partitioning of
time, or equivalently bandwidth, between pilots and data within a
coherence interval was also optimally selected. We found that,
with $100$ antennas at the BS, by optimally allocating energy to
pilots, the energy efficiency can be increased as much as $50\%$,
when each terminal has a throughput of about $1$ bit/s/Hz.
Typically, when the SNR is low (e.g., around $-15$dB), at the
optimum, the transmit power is then about $10$ times higher during
the training phase than during the data transmission phase.

\appendix

\subsection{Proof of Proposition~\ref{lemma 1}} \label{proof Lemm1}
From \eqref{eq: low SE 1} and \eqref{Prob2}, the problem
$\mathcal{P}_2$ becomes
\begin{align}
    \mathcal{P}_2 = \left\{
\begin{array}{l}
  \arg \mathop {\max}\limits_{\Pu} ~ \left(1-\frac{K}{T}\right)
    \sum_{k=1}^K
        \log_2
        \left(
            1
            +
            f_k\left(\Pu\right)
        \right)\\
  \hspace{1.7cm} 0 \leq \Pu \leq  \frac{P}{T-K}\\
\end{array}
\right.
\end{align}
where
\begin{align*}
    &f_k\left(\Pu\right)
    \nonumber
    \\
    &\triangleq
            \frac{
                a_k \left(P-\left(T-K\right)\Pu \right)\Pu
                }{
                b_k \left(P\!-\!\left(T\!-\!K\right)\Pu \right)\Pu + c_k \Pu + d_k \left(P\!-\!\left(T\!-\!K\right)\Pu \right) +1
                }
    \\
    &=\!
    \frac{a_k}{b_k} \!-\!
    \frac{a_k}{b_k}
    \frac{c_k \Pu + d_k \left(P-\left(T-K\right)\Pu \right) +1}
        {b_k \!\left(\!P\!\!-\!\left(\!T\!\!-\!K\!\right)\Pu \!\right)\Pu \!+\! c_k \Pu \!+\! d_k \!\left(\!P\!\!-\!\left(\!T\!\!-\!K\!\right)\Pu \!\right)
    \!+\!1}.
\end{align*}
The second derivative of $f_k\left(\Pu\right)$ can be expressed
as:
\begin{align} \label{eq: prop2 1}
&\omega_k\frac{\partial^2 f_k\left(\Pu\right)}{\partial \Pu^2} =
-b_k\hat{T}^2(c_k-d_k\hat{T})\Pu^3 -
3b_k\hat{T}^2(d_kP+1)\Pu^2\nonumber\\
&+3b_k\hat{T}P(d_kP+1)\Pu - (d_kP+1)(b_kP^2+c_kP+\hat{T}),
\end{align}
where $\omega_k\triangleq \frac{\left( b_k \left(P\!-\!\hat{T}\Pu
\right)\Pu + c_k \Pu + d_k \left(P\!-\!\hat{T}\Pu \right)
+1\right)^3}{2a_k}$, and $\hat{T}\triangleq T-K$. Since $P\geq
\hat{T}\Pu$, we have
\begin{align} \label{eq: prop2 1}
&\omega_k\frac{\partial^2 f_k\left(\Pu\right)}{\partial \Pu^2} =
-b_kc_k\hat{T}^2\Pu^3 -(d_kP+1)(c_kP+\hat{T})\nonumber\\
&-\frac{3}{4}b_k\hat{T}^2\Pu^2-b_k\left(P-\frac{3}{2}\hat{T}\Pu\right)^2-b_kd_k(P-\hat{T}\Pu)^3\leq
0.
\end{align}
Since $\omega_k>0$, $\frac{\partial^2
f_k\left(\Pu\right)}{\partial \Pu^2}\leq0$.
Therefore, $f_k\left(\Pu\right)$ is a concave function in $0\leq
\Pu \leq \frac{P}{T-K}$. Since $\log_2\left(1+x\right)$ is a
concave and increasing function,
$\log_2\left(1+f_k\left(\Pu\right)\right)$ is also a concave
function. Finally, using the fact that the summation of concave
functions is concave, we conclude the proof of
Proposition~\ref{lemma 1}.

%\bibliographystyle{IEEEtran}
%\bibliography{IEEEabrv,CCTLABBiblio}


\end{document}